# Sedimentation of rapidly interacting multicomponent systems

P. Schuck*, S.K. Chaturvedi

Dynamics of Macromolecular Assembly Section, Laboratory of Cellular Imaging and Macromolecular Biophysics, National Institute of Biomedical Imaging and Bioengineering, National Institutes of Health, United States

*schuckp@mail.nih.gov

**Abstract**

The biophysical analysis of dynamically formed multi-protein complexes in solution presents a formidable technical challenge. Sedimentation velocity (SV) analytical ultracentrifugation achieves strongly size-dependent hydrodynamic resolution of different size species, and can be combined with multi-component detection by exploiting different spectral properties or temporally modulated signals from photoswitchable proteins. Coexisting complexes arising from self- or hetero-associations that can be distinguished in SV allow measurement of their stoichiometry, affinity, and cooperativity. However, assemblies that are short-lived on the time-scale of sedimentation ($t_{1/2}$ < 100 sec) will exhibit an as of yet unexplored pattern of sedimentation boundaries governed by coupled co-migration of the entire system. Here, we present a theory for multi-component sedimentation of rapidly interacting systems, which reveals simple underlying physical principles, offers a quantitative framework for analysis, thereby extending the dynamic range of SV for studying multi-component interactions.

**Introduction**

Reversible self-association and hetero-association processes are a key aspect of macromolecular function, and play a central role in cellular processes and in the assembly of cellular structures across a spectrum of architectures, from stable and well-defined supramolecular entities (1), to polymorph and transient signaling particles (2, 3), or liquid-liquid phase-separated cellular bodies (4). Hallmarks are multi-valency and cooperativity promoting the formation of multi-protein complexes (5). Thus, methods for studying protein interactions with regard to their assembly states and energetics in multi-component solutions are important.

Sedimentation velocity analytical ultracentrifugation (SV) is a first-principles-based classical transport method well-suited to study protein interactions, taking advantage of the strongly mass-and-shape-dependent migration in the centrifugal field to unravel populations of different co-existing species free in solution (6, 7). In the last decades, the method has significantly advanced in resolution and sensitivity with the commercial availability of a fluorescence detector (8), in combination with modern computational methods to calculate diffusion-deconvoluted sedimentation coefficient distributions $c(s)$ with high hydrodynamic resolution from noisy data (9). For the analysis of multi-component protein complexes, SV was further enhanced with multi-signal and multi-wavelength analysis for conventional detection systems used in the μM range (10), and, as shown in our recent work, by monochromatic multicomponent analysis for high-affinity interactions detected with time-dependent fluorescence signals (11) in the nanomolar range and below. These additional spectral/temporal signal dimensions provide independent information on the composition of each resolved sedimenting species, thereby facilitating the identification of the complexes formed by mass (or sedimentation coefficient) and composition. For example, these techniques allowed the analysis of competitive high-affinity homo-and-heterodimer formation of glutamate receptor domains, thought to dictate the properties of synaptic transmission (11–13), and the study of ternary adaptor protein complexes leading to intracellular receptor crosslinking, facilitating the formation of signaling complexes in T-cell activation (2, 14).

However, for interacting systems where complexes are formed but do not persist for the duration of the sedimentation experiment (>1000 sec), more complicated sedimentation processes emerge, as the complexes sediment at a higher rate than their free constituents they rapidly exchange with. Furthermore, due to the geometry of SV, complexes remain in a bath of their constituents, unlike, for example, in chromatography, where interacting components will separate. The resulting `reaction boundaries', recognized by concentration-dependent boundary velocities, have been off-limits for direct interpretation since the discovery of reversible protein interactions in the 1930s (6). Seminal studies of the asymptotic properties of sedimentation boundaries in rapidly equilibrating two-component interacting mixtures by (15) have confirmed their mysterious behavior rather than clarified their mechanism: This includes the fact that boundary velocities do not reflect that of any sedimenting species, may show increasing or decreasing concentration-dependence, are usually --- but not always --- accompanied by `undisturbed' boundaries of one of the free species, and that patterns may discontinuously change at component loading concentration ratios that do not correspond to the complex stoichiometry (16). In the context of fluorescence-detected SV, we recently discovered additional anomalies in the measured transport behavior in the case of quantum yield changes associated with complex formation (17).

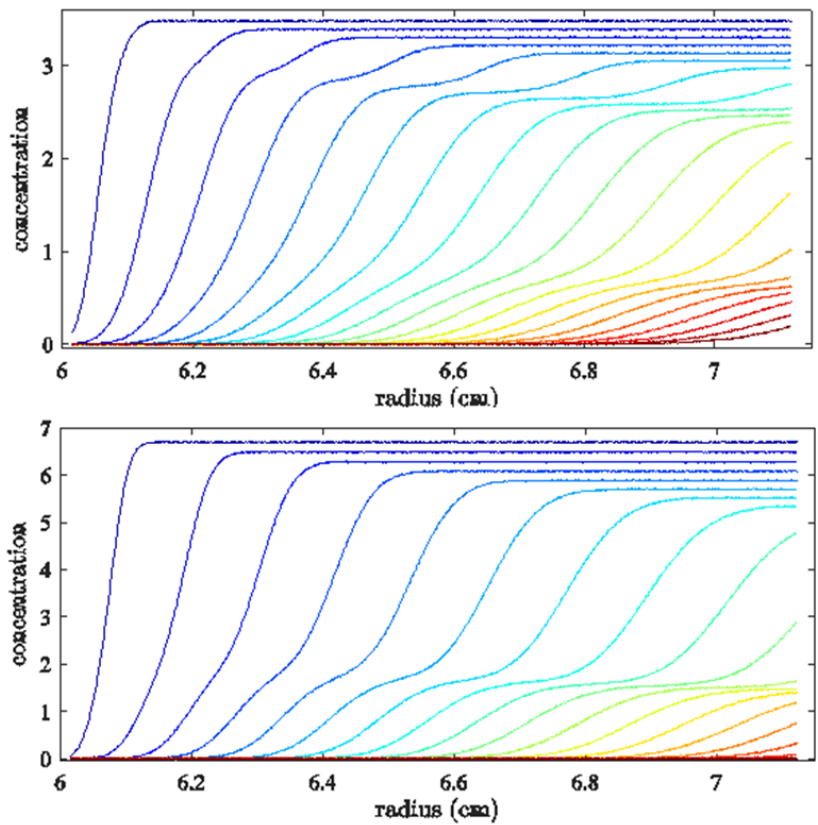

**Figure 1**: Evolution of sedimentation profiles for a three-component system describing a molecule *A* with independent sites for *B* and *C*, equilibrating rapidly on the time-scale of sedimentation. It was assumed that *A* and *C* each have molar masses of 150 kDa and *B* is 100 kDa, with equilibrium dissociation constants $K_{AB} = K_{AC} = 1$ μM, and *s*-values of $s_A = 7$ S, $s_B = 5.5$ S, $s_C = 8$ S, $s_{AB} = 10$ S, $s_{AC} = 11.5$ S, and $s_{ABC} = 15$ S. Eq. 3 was solved for a 12 mm solution column at 50,000 rpm, and total signals assuming mass-based signal increments are plotted in 600 sec intervals, plotted in increasing color temperature at later times (with equivalent colors for both plots).

To illustrate the problem, Fig. 1 shows sedimentation patterns of an ensemble of molecules *A* in a mixture with two different ligands *B* and *C*. Such a situation is readily encountered in practice, for example, when studying bi-specific antibodies. In fact, mutually mono-valent ternary systems may be considered a minimal biologically meaningful interaction pattern. The partial-differential equation (PDE) of the coupled sedimentation/reaction/diffusion process, the Lamm equation Eq. 3 (16) is the master equation of SV and can predict the evolution of concentration profiles for all species, but it has no explanatory power for the resulting features: Why does sedimentation occur in distinct boundaries? Why can we sometimes recognize two and sometimes three boundaries? What is their composition? Why are their *s*-values different and concentration-dependent? How do they reflect on the sedimenting molecules and the binding constants? Perhaps surprisingly, despite the long history of SV, these questions currently cannot be answered. They are pertinent, in particular, since coupled Lamm PDEs cannot be universally applied for data analysis due to their high susceptibility to sample imperfections (18), whereas boundaries can usually readily be recognized and quantified.

For two-component systems, effective particle theory (EPT) was developed to address such questions, and it naturally explains the features of sedimentation boundaries of rapidly interacting systems by considering as the sedimenting `species' only the subset of dynamically interconverting free species and complexes that exhibit coupled co-sedimentation due to matching constituent time-average velocities (19). Diffusional spread of boundaries can be predicted in the same framework (20). This physically meaningful picture can correctly predict the concentration-dependence of the most robust observables, which are the boundary amplitudes, $s$-values, and composition, and allows quantitative analysis of experimental data (21, 22). In view of the need to consider interactions between more than two macromolecular components, matching current detection capabilities (10, 11), here we describe a more general framework for effective sedimenting particles in a solution of an arbitrary number of interacting components and complexes, which can explain, for example, the principles governing sedimentation boundaries such as those in Fig. 1.

**<u>Theory</u>**

We consider a mixture of $K$ macromolecular components, enumerated as $\kappa = 1 \ldots K$, that can form $E$ complexes $e = 1 \ldots E$, such that there are a total of $N = K + E$ species in solution. The species are enumerated with the index $i = 1 \ldots N$, and ordered such that $i \leq K$ refer to the free species of the components, and species indices $i > K$ are the complexes. We express the linked equilibria in an unambiguous way by considering, for each complex, only the reaction pathway between the complex and the free species of the constituent macromolecular components of that complex. This leads to a set of $E$ reactions, one for each complex, of the form

$$n_{1,e}[\text{component 1}] + n_{2,e}[\text{component 2}] + \ldots + n_{K,e}[\text{component } K] \leftrightarrow [\text{complex } e] \quad \text{(Eq. 1)}$$

where the brackets indicate concentrations of free species, and $n_{\kappa,e}$ are the number of copies of molecule $\kappa$ participating in the complex $e$.

We assume that mass action law holds locally at all times, such that the concentration of complex $e$ (species $i = K + e$) is

$$\chi_{K+e} = \exp\left(-\frac{\Delta G_e}{RT}\right) \prod_{i=1}^{K} (\chi_i)^{n_{i,e}} \quad \text{(Eq. 2)}$$

The sedimentation of each species is described by the Lamm equation (16)

$$\frac{\partial \chi_i}{\partial t} + \frac{1}{r}\frac{\partial}{\partial r}\left(\chi_i s_i \omega^2 r^2 - D_i \frac{\partial \chi_i}{\partial r} r\right) = q_i \quad \text{(Eq. 3)}$$

with $s_i$ and $D_i$ denoting the species' sedimentation and diffusion coefficient, $\omega$ the rotor angular velocity, $r$ the radial position, and $t$ time. It describes the local balance between transport fluxes and the net chemical reaction flux $q_i$ maintaining quasi-instantaneously mass action law Eq. 2 according to the local concentrations of all species.

Because all macromolecules are considered in instantaneous exchange with their complexes, a key quantity is the time-average velocity experienced by each, or the constituent velocity

$$\overline{s_\kappa} = \frac{1}{\chi_{\kappa,\text{tot}}}\left(\chi_\kappa s_\kappa + \sum_{e=1}^{E} n_{\kappa,e}\chi_{K+e}s_{K+e}\right) \tag{Eq. 4}$$

with

$$\chi_{\kappa,\text{tot}} = \chi_\kappa + \sum_{e=1}^{E} n_{\kappa,e}\chi_{K+e} \tag{Eq. 5}$$

denoting the total component concentration.

It is obvious that the entire system cannot sediment as a single boundary, since (except for special cases discussed below) different components exhibit different time-average velocities $\overline{s_\kappa}$. Further, because attainment of chemical equilibrium is assumed to be quasi instantaneous, nowhere in the solution can complexes be separated from their free constituent components. Given the geometry of sedimentation with initially uniform solutions, it immediately follows that there will be at most $K$ boundaries separating as many zones in the solution. The first one will consist of a single-component solution, the second of a two-component mixture, and so on, up to a $K$-component mixture in the last (and fastest) boundary that has a composition identical to the loading mixture. As sketched in Fig. 2, the components in each zone will be in different chemical equilibrium, and the concentration difference between neighboring zones will be observed to co-sediment with the separating sedimentation boundary.

EPT is based on the time-honored approximation of a rectangular solution column with constant force (15, 19, 23). With the focus on the boundary pattern and their associated average transport, as opposed to boundary shapes, diffusion is neglected. Based on the transport method, this leaves the average $s$-value of a boundary invariant, and is commensurate with quantities measured by integration of diffusion-deconvoluted sedimentation coefficient distributions $c(s)$ (7, 9, 24). This motivates solutions of the coupled Lamm equations with a sum of Heaviside step-functions describing series of moving boundaries

$$\chi_i(r,t) = \sum_{b=1}^{K}\left(\chi_{i,b} - \chi_{i,b-1}\right)H(r - s_b t) \tag{Eq. 6}$$

(where $s_b$ is the velocity of boundary $b$) with amplitudes corresponding to the species' concentration differences across the associated zones. (Of historical interest, generalized functions such as step-functions and Dirac's $\delta$-functions were not readily available at the time of Gilbert (25).) As described in the **Supplementary Theory**, Eq. 6 leads to a set of algebraic rules for the boundary amplitudes and velocities that have direct physical meaning.

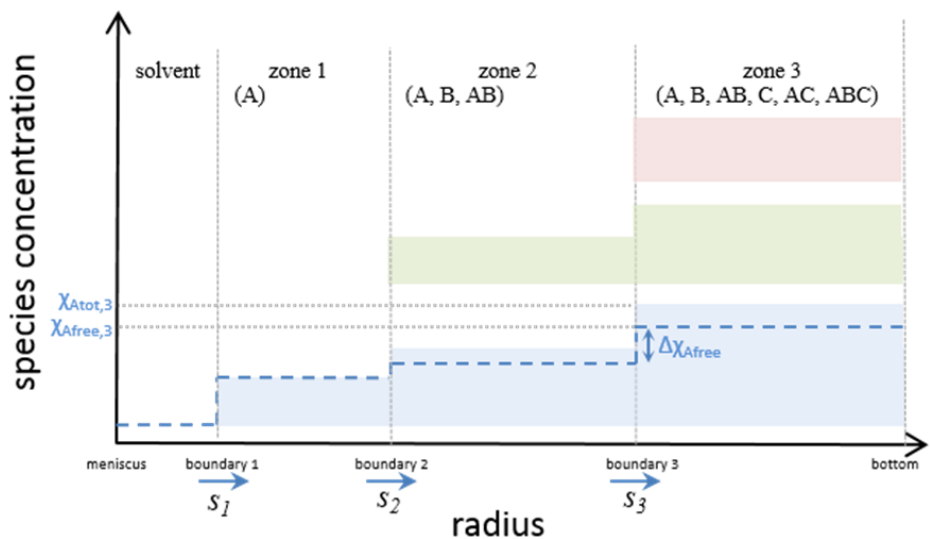

**Figure 2**: Schematics of the boundary pattern for a three-component system depicted in the particular configuration where *A* (blue) provides the slowest `undisturbed' boundary, *B* (green) is dominant component in the middle reaction boundary, and *C* (red) is dominant and exclusive in the fastest reaction boundary. After the solution column is initially uniformly loaded at the start of sedimentation, the boundaries move with the velocities $s_1$, $s_2$, and $s_3$, respectively, forming different zones in the solution represented here at a particular point in time. The boundary velocities are determined by the macromolecular composition in the respective zones, illustrated here with a system where molecule *A* has independent sites for *B* and *C*. The ordinate depicts a relative concentration scale, with the total component concentrations indicated as colored patches, offset for clarity, and with the height of the patches reflecting the total component concentration in each zone. Molecules of each component will assume states of all possible species in each zone, dependent on other components in the same zone as well as the equilibrium constants. As a particular example of a species, highlighted as blue dashed line is the concentration of the free form of *A*. The difference in species concentration to the next higher zone (e.g., $\Delta\chi_{Afree,3}$) is recognized to be co-sedimenting with the boundary separating the zones, to maintain identical time-average $s$-values of molecules from all components co-sedimenting in each zone.

## **Results**

The boundary patterns of rapidly reversible interacting systems are fully described by the following principles: (1) In any given zone, the component with the highest time-average velocity ($\overline{s_\kappa}$) will be completely engulfed in that zone and determine $s$-value of the associated boundary; it is therefore referred to as `dominant' in that boundary. (2) The co-sedimenting fractions of all other components pulled along will be of a magnitude such that their time-average velocity matches that of the dominant component. This allows for dynamic co-sedimentation of all species migrating in that boundary. (3) What is left behind establishes the composition of the next lower zone, which assumes a new chemical equilibrium.

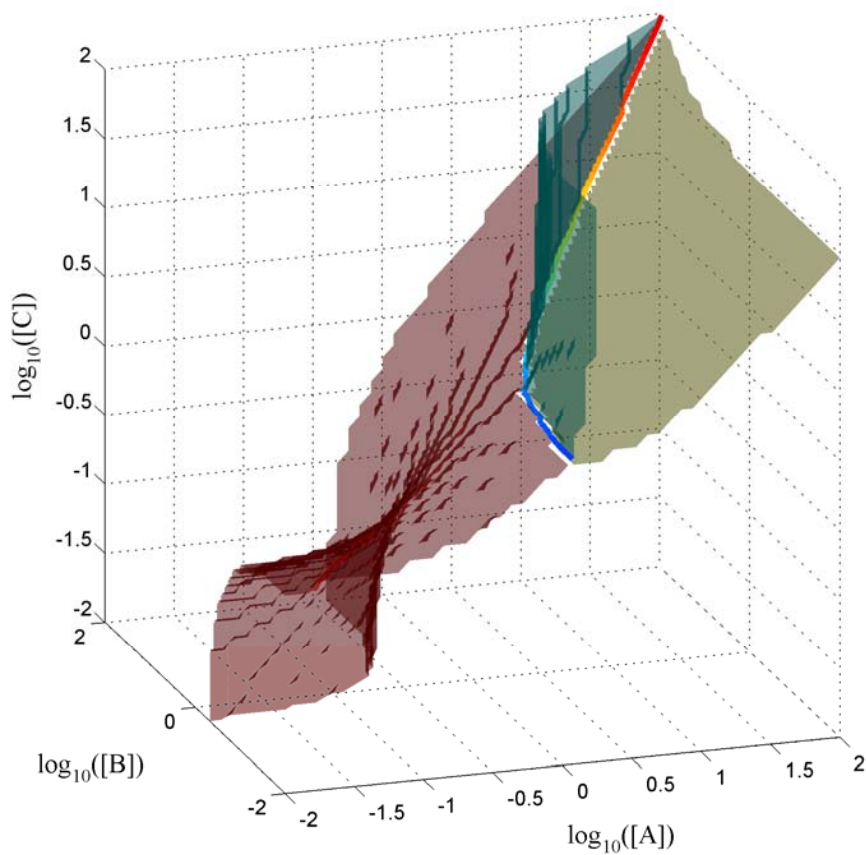

**Figure 3**: For the three-component, two-site system of Fig. 1, the 'phase transitions' where the dominant component of the fastest reaction boundary switches (**A** / **B** transition in basil; **A** / **C** in maroon; **B** / **C** in olive). These surfaces meet in the line describing loading concentrations where the entire system sediments in a single boundary, at an $s$-value indicated by color temperature (scale in Supplementary Fig. S1).

These simple rules create remarkably rich boundary patterns in their dependence on the total loading concentrations of all components. They can be classified according to the combination of components in each zone. When two components have the same time-average velocity $\overline{s_\kappa}$ in one zone, they are both excluded from the lower zones and one fewer sedimentation boundary occurs. Any higher concentrations of one or the other component will cause the lower zones to be composed of different combination of components, which will cause discontinuities in sedimentation behavior of the lower boundaries as a function of concentration (`phase transitions'). Conditions of matching $\overline{s_\kappa}$ between pairs of components define $(K-1)$-dimensional hyper-surfaces in concentration space, and all hypersurfaces meet in a single line in concentration space that describes concentrations where the entire system sediments in a single boundary (Fig. 3).

As a test of the quality of predictions by EPT, we revisit the system of Fig. 1 where *A* has independent sites for *B* and *C*, with *B* and *C* non-interacting, such as a bi-specific antibody interacting with its two independent ligands. Using the full Lamm PDE Eq. 3, we calculated sedimentation boundaries as in Fig. 1 and from these the corresponding $c(s)$ sedimentation coefficient distributions representing the boundary pattern Fig. 4. We used loading concentrations along different titration series of constant *A* and *C*, or constant *B* and *C*, respectively. As shown in the *Left* panel of Fig. 4, the titration series of *B* produces three boundaries, two of which remain at constant $s$-value coincident with those of two free species but change in height (circles and triangles), and the fastest boundary (squares) at monotonically increasing $s$-values (not coinciding with any species). The titration series of *A* (*Right Panel*) exhibits a more complicated pattern, with slow boundary (circles) discontinuously changing position among the free species $s$-values, the fastest reaction boundary $s$-value (squares) decreasing then increasing then decreasing again as a function of concentration, and a middle boundary (triangles) that is sometimes coincident with a free species, sometimes a secondary reaction boundary. Throughout, EPT and $c(s)$ from the full Lamm PDE are in excellent agreement. Supplementary Figs. S1 and S2 show a representation of the fastest and the medium reaction boundary in three-dimensional concentration space.

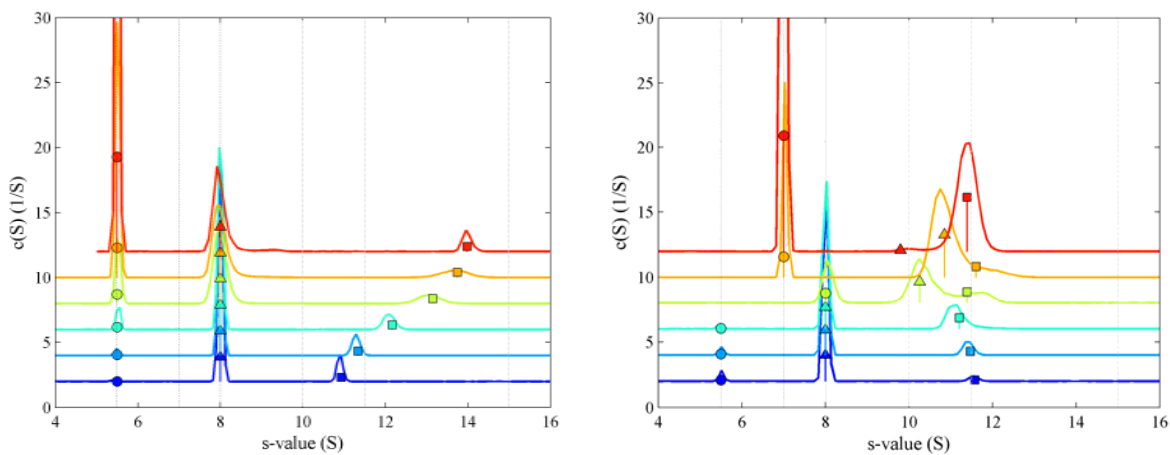

**Figure 4**: For the same system as in Fig. 1, a comparison between boundary $s$-values and amplitudes obtained from sedimentation coefficient distribution modeling of a full simulation of the Lamm PDE Eq. 3 (solid lines, arbitrarily offset for clarity) and the predictions from effective particle theory for the undisturbed boundary (circles), the middle reaction boundary (triangles), and the fastest reaction boundary (squares). Vertical grey dotted lines depict $s$-values of free and complex species. *Left:* Total loading concentrations were along a titration series of constant 0.36 μM *A* and constant 5.0 μM *C* with increasing *B* at 0.095, 0.29, 0.91, 2.8, 8.7, and 27 μM (higher concentrations indicated by increasing color temperature). *Right*: Titration series of constant 0.36 μM *B* and constant 5.0 μM *C* with increasing *A* at 0.095, 0.29, 0.91, 2.8, 8.7, and 27 μM. For comparison of boundary amplitudes, integral representation of the same sedimentation coefficient distribution $c(s)$ and predicted by EPT can be found in Fig. 5.

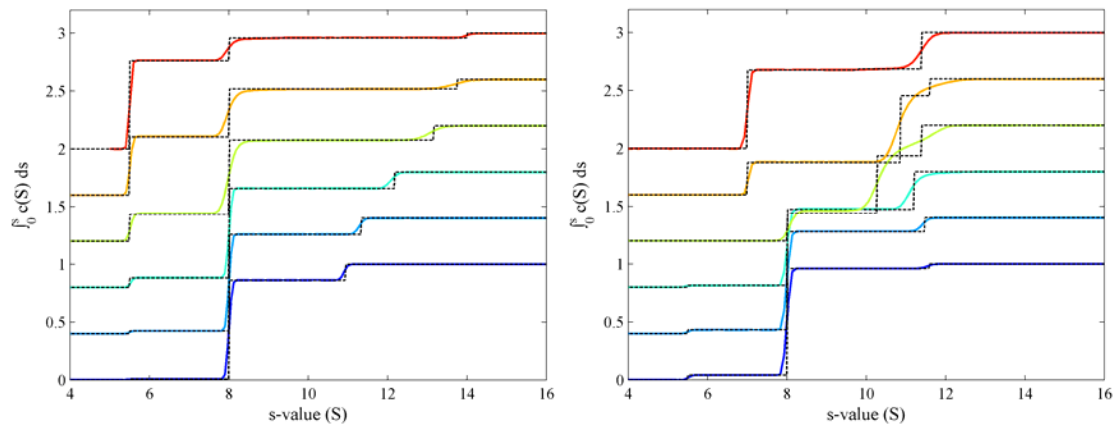

**Figure 5**: Corresponding to Fig. 4, integral representations of the same distributions. *Left*: Titration series of constant ***A*** and ***C*** with increasing ***B***. *Right*: Titration series of constant ***B*** and constant ***C*** with increasing ***A***.

## Discussion

Even though many high-affinity interactions will exhibit sufficiently long complex life-times for the hydrodynamic separation of interacting species, this is not necessarily always the case. Therefore, a framework for the interpretation of sedimentation coefficient distributions for rapidly reversible systems is important, and will aid also in the study of moderate affinity systems with conventional detectors.

The ability to predict the sedimentation boundary patterns of rapidly reversible systems addresses a fundamental problem encountered in the study of such systems, which show richly featured sedimentation profiles but with amplitudes and $s$-values not corresponding those of the interacting species. The dynamically coupled co-sedimentation process can be fit with Lamm equation PDEs, but in practice these are overly sensitive to all factors governing the boundary shapes, besides diffusion and detailed kinetic rate constants also including micro-heterogeneity (e.g., from chemical modifications, or differential glycosylation) and other sample imperfections as well as hydrodynamic non-ideality (18, 21). We have shown previously how EPT allows interpretation of the more robust features of the observed sedimentation --- the number, velocities, and amplitudes of the boundaries --- which can be easily extracted from the sedimentation coefficient distributions (21). The present work allows the same strategies to be used for multi-component systems.

The extension of EPT from two-components to an arbitrary number of components and complexes solves a long-standing conceptual problem in analytical ultracentrifugation arising from the limitation of the Lamm PDE to predict but not explain the boundary patterns of dynamically coupled sedimentation processes. It is satisfactory that the simple principles of matching constituent time-average sedimentation velocity hold for the general case, and fully explain the observed phenomenology.

*Acknowledgements*

This work was supported by the Intramural Research Program of the National Institute of Biomedical Imaging and Bioengineering at the National Institutes of Health, United States.

**Supplementary Theory**

*Rectangular Lamm equation without diffusion*

EPT is based on the approximation of a rectangular solution column with constant force (15, 19, 23), eliminating complications from the radial geometry. Also, as Gilbert-Jenkins theory, it is neglecting diffusion (15). This simplifies the Lamm PDE Eq. 3 to

$$\frac{\partial \chi_i}{\partial t} + s_i \frac{\partial \chi_i}{\partial r} = q_i \tag{Eq. S1}$$

The neglect of diffusion restricts the analysis to the overall transport in each boundary, which does not depend on boundary shapes and may be described by a step-function at the second-moment position of the boundary (24).

*Reaction fluxes*

The reaction fluxes $q_i$ in Eq. Eq. 3 and S1 account for the sum of all reactions $e$ potentially involving the formation or dissociation of species $i$

$$q_i = \sum_{e=1}^{E} q_{i,e} \tag{Eq. S2}$$

It is convenient to extend the stoichiometries $n_{\kappa,e}$, defined in Eq. 1 above as the number of copies of component $\kappa = 1\ldots K$ contributing to complex $e$, to all species and to define $n_{i>K,e} = -\delta_{i-K,e}$ (with the Kronecker symbol producing values of $-1$ for $n_{e+K,e}$ and $0$ otherwise), such that reactants and products of a reaction can be accounted for in the same way: With this notation the $q_{i,e}$ take the general form

$$q_{i,e} = -n_{i,e}\left(k_{\text{on},e}\prod_{\kappa=1}^{K}\chi_\kappa^{n_{\kappa,e}} - k_{\text{off},e}\chi_e\right) =: -n_{i,e} r_e \tag{Eq. S3}$$

describing changes in the molar species concentrations arising from binding and dissociation reactions. The dissociation rate constants $k_{\text{off},e}$ are high, for fast local equilibration, but unknown, and so are the $r_e$. However, we can still express the total reaction fluxes $q_i$ as

$$q_i = -\sum_{e=1}^{E} n_{i,e} r_e \tag{Eq. S4}$$

In matrix/vector notation for all species, this is

$$\vec{q} = \mathbf{S}\vec{r} \tag{Eq. S5}$$

with the ($N \times E$) matrix $\mathbf{S}$ composed of the stoichiometries $S_{i,e} = -n_{i,e}$ for each reaction, and the ($E \times 1$) column vector $\vec{r}$ containing the unknown reaction rates $r_e$ for each reaction.

An effective strategy to eliminate the unknown reaction fluxes is to find linear combinations such that they cancel out. Thus we are looking for sets of coefficients $A_{i,\kappa}$ such that

$$\sum_{i=1}^{N} q_i A_{i,\kappa} = 0 \qquad \text{(Eq. S6)}$$

Therefore, we seek a matrix $\mathbf{A}$ such that $0 = \mathbf{A}^T \vec{q}$, or with Eq. S5 equivalently $0 = \mathbf{A}^T \mathbf{S}\vec{r}$, which leads to the requirement

$$\mathbf{S}^T \mathbf{A} = 0 \qquad \text{(Eq. S7)}$$

Singular value decomposition $\mathbf{S} = \mathbf{U}\boldsymbol{\Sigma}\mathbf{V}$, leads to a ($N \times N$) matrix $\mathbf{U}$ with columns $E+1\ldots N$ spanning a basis of the null-space of $\mathbf{S}$, which is of dimension $N - E = K$ (26). Thus, the choice $A_{i,\kappa} = U_{i,E+\kappa}$ satisfies Eq. S6. However, we can find a more useful choice of coefficients by joining the identity matrix and the stoichiometries:

$$A_{i,\kappa} = \begin{pmatrix} 1 & & 0 \\ & 1 & \\ 0 & & 1 \\ n_{11} & \cdots & n_{K1} \\ \vdots & & \vdots \\ n_{1E} & \cdots & n_{KE} \end{pmatrix} \qquad \text{(Eq. S8)}$$

can be readily shown to satisfy Eq. S7. Therefore, we can make use of $A_{i,\kappa}$ to eliminate reaction fluxes <u>via</u> Eq. S6. The structure of $A_{i,\kappa}$ is such that it has non-zero entries in a column $k$ only for species containing component $\kappa$. Therefore, the total constituent concentration (in protomer units) of any component $k$ can be conveniently calculated as

$$\bar{\chi}_k = \sum_{i=1}^{N} \chi_i A_{i,k} \qquad \text{(Eq. S9)}$$

and the constituent velocity $\bar{s}_\kappa$ from Eq. 4 as

$$\bar{s}_k = \frac{\displaystyle\sum_{i=1}^{N} \chi_i s_i A_{i,k}}{\displaystyle\sum_{i=1}^{N} \chi_i A_{i,k}} \qquad \text{(Eq. S10)}$$

Another physical interpretation of $A_{i,\kappa}$ is that the products $\chi_i A_{i,\kappa}$ reflect the relative time a molecule from component $\kappa$ spends in the state of species $i$ (after normalization relative to total concentration of $\kappa$).

*Coupled transport*

We now insert Eq. 6 into S1. The spatial and temporal derivative turns the Heaviside step-functions into Dirac's delta-functions (25)

$$\frac{\partial \chi_i(r,t)}{\partial t} = \sum_{b=1}^{K} (\chi_{i,b} - \chi_{i,b-1})(-s_b)\delta(r - s_b t)$$

$$\frac{\partial \chi_i(r,t)}{\partial r} = \sum_{b=1}^{K} (\chi_{i,b} - \chi_{i,b-1})\delta(r - s_b t)$$

(Eq. S11)

and we find

$$\sum_{b=1}^{K} (\chi_{i,b} - \chi_{i,b-1})(s_i - s_b)\delta(r - s_b t) = q_i \quad \forall i = 1...N$$

(Eq. S12)

This can be inserted into Eq. S6 and we observe

$$\sum_{i=1}^{N} \sum_{b=1}^{K} (\chi_{i,b} - \chi_{i,b-1})(s_i - s_b)\delta(r - s_b t) A_{i,\kappa} = 0 \quad \forall \kappa$$

(Eq. S13)

Since this must be true for all $r$ and $t$, it must hold for all boundaries separately

$$\sum_{i=1}^{N} (\chi_{i,b} - \chi_{i,b-1})(s_i - s_b) A_{i,\kappa} = 0 \quad \forall b, \kappa$$

(Eq. S14)

and rearranging, we find the boundary velocities

$$s_b = \frac{\sum_{i=1}^{N} (\chi_{i,b} - \chi_{i,b-1}) s_i A_{i,\kappa}}{\sum_{i=1}^{N} (\chi_{i,b} - \chi_{i,b-1}) A_{i,\kappa}} \quad \forall b, \kappa$$

(Eq. S15)

which this is true for any choice of $\kappa$. As noted above, the products $\chi_i A_{i,\kappa}$ reflect the relative time a molecule from component $\kappa$ spends in the state of species $i$. Therefore, it follows from Eq. (S15) that any molecules contributing to the concentration steps in a boundary, i.e., all species 'co-sedimenting' with a boundary (including $\Delta\chi_{\text{Afree}}$ in Fig. 2), exhibit the same time-average velocity $s_b$.

We can express Eq. (S14) with the help of the constituent concentrations of component $k$ in a zone $b$, $\overline{\chi}_{k,b}$, and the corresponding constituent velocities $\overline{s_{k,b}}$ and find

$$(s_b - \overline{s_{k,b}})\overline{\chi}_{k,b} = (s_b - \overline{s_{k,b-1}})\overline{\chi}_{k,b-1}$$

(Eq. S16)

We can interpret this to be a continuity in the flux density of molecules $k$ falling behind the moving boundary (if their velocity in zone $b$ is lower than that of the moving boundary), and emanating from it in zone $b-1$ migrating there both at lower concentration and lower velocity.

*Boundary structure*

The calculation of the boundary properties can proceed iteratively from the fastest to slowest boundary. Starting point is the loading composition, which --- in the absence of radial dilution --- equals the composition of the highest zone $K$

$$\chi_{i,K} = \chi_{i,\text{load}} \tag{Eq. S17}$$

For principal reasons, one component must be contained exclusively in this zone, i.e., be 'dominant' for the highest boundary (Fig. 2). Let us label this one as $d_K$. Since it is exclusively in zone $K$, there is no free species $d_K$ in the lower boundaries, $\chi_{d_K,b<K} = 0$, and therefore all complexes to which it contributes vanish in the lower boundaries. Since our particular choice of $A_{i,\kappa}$ has non-zero elements only for species $i$ that contain a component $\kappa$, it follows $\sum_i \chi_{i,K-1} A_{i,d_K}$ vanishes, and so does $\sum_i \chi_{i,K-1} s_i A_{i,d_K}$. Therefore, using Eq. (S15) for the index $\kappa = d_K$ we find the velocity of the highest boundary to be

$$s_K = \frac{\sum_{i=1}^{N} \chi_{i,\text{load}} s_i A_{i,d_K}}{\sum_{i=1}^{N} \chi_{i,\text{load}} A_{i,d_K}} = \overline{s_{d_K}} \tag{Eq. S18}$$

i.e., identical to the time-average (or constituent average) velocity of the dominant component. Which component is the dominant one can be found through inspection of constituent quantities in Eq. (S16) for $b = K$. Considering that there can only be positive concentrations, and that $s_K$ must be faster than any constituent velocity in the boundary below, the left-hand side of Eq. (S16) must not be negative. As a consequence, $s_K$ must be faster or equal to any constituent velocity. The answer is therefore straightforward and we can identify $d_K$ with the component that exhibits the highest time-average velocity $\overline{s_{K,b}}$.

Finally, we can now calculate the co-sedimenting species from all other components, by rephrasing Eq. (S14) to solve for the $\chi_{i,b-1}$ as

$$\sum_{i=1}^{N} \chi_{i,K-1}(s_K - s_i) A_{i,\kappa} = \sum_{i=1}^{N} \chi_{i,\text{load}}(s_K - s_i) A_{i,\kappa} \quad \forall \, \kappa \neq d_K \tag{Eq. S19}$$

where the right-hand-side is known. Since in the zone $K-1$ all complex species concentrations are linked to the free concentrations by mass action law, this is a system of $K-1$ equations for the $K-1$ unknown free concentrations $\chi_{\kappa \neq d_K, K-1}$ of components in the lower zone, from which all other species in the lower boundary are automatically determined through mass action law Eq. 2.

For the next lower zone, knowing the component composition of the zone allows us to recursively apply the same considerations as for the highest boundary, until we reach the lowest zone that consists of only a single component. Unless that component exhibits self-association, it will be a concentration-independent `undisturbed' boundary.

*Combinatorics of boundary compositions and `phase transitions'*

The algorithm above provides a recipe for determining the combination of components that are present in each zone, i.e., the sequence of components $d_b$ that are dominant for the sedimentation boundary of the sequential zones. For a specific interacting system with given species $s$-values and binding constants, this is unambiguously determined by the $K$ loading concentrations.

There can be concentrations where a non-dominant component that has the same constituent-average velocity as the dominant component. From Eq. (S16) we can see that for any non-dominant component $x$ with

$$\overline{s_{x,b}} = \overline{s_b} = \overline{s_{d_b,b}} \tag{Eq. S20}$$

and Eq. (S15) shows that the concentrations in the next lower boundary $\overline{\chi}_{x,b-1}$ vanish, meaning that this component is also entirely contained in the zone $b$ co-sedimenting with the associated boundary, and fewer (or no) lower boundaries exist.

In concentration space, the condition Eq. (S20) defines a transition point: any lower concentration of $x$ will decrease $\overline{s_{x,b}}$ below $\overline{s_b}$ and create fractions not co-sedimenting, and any higher concentrations of $x$ will increase $\overline{s_{x,b}}$ and make it the new dominant component. Due to the change in propagation mode and the associated switch in the nature of the lower boundaries (e.g., often changing the nature undisturbed boundary of free species) we refer to this as a `phase transition'. All properties change continuously across this point in concentration space, though not the concentration first derivatives (such as slopes in isotherms).

There can be phase transitions for all reaction boundaries, including the highest boundary as well as the lower ones. For the phase transitions of the highest boundary, for two-component systems Eq. (S20) defines a phase transition line (19). For three-component systems it defines a surface due to the additional dependence of species populations on the third component $y$ with $y \neq x$ and $y \neq d_b$.

The dimensionality of the concentration space that satisfies Eq. (S20) for any component is reduced by 1. If two components satisfy Eq. (S20) simultaneously, it will be reduced by 2, and so on. Finally, if all (non-dominant) components satisfy Eq. (S20), the phase transitions will meet in a line that defines points in concentration space for which there is only a single boundary for the entire system. This line will be where the phase transition surfaces meet. An example can be found in Fig. 3.

**Supplementary Figures**

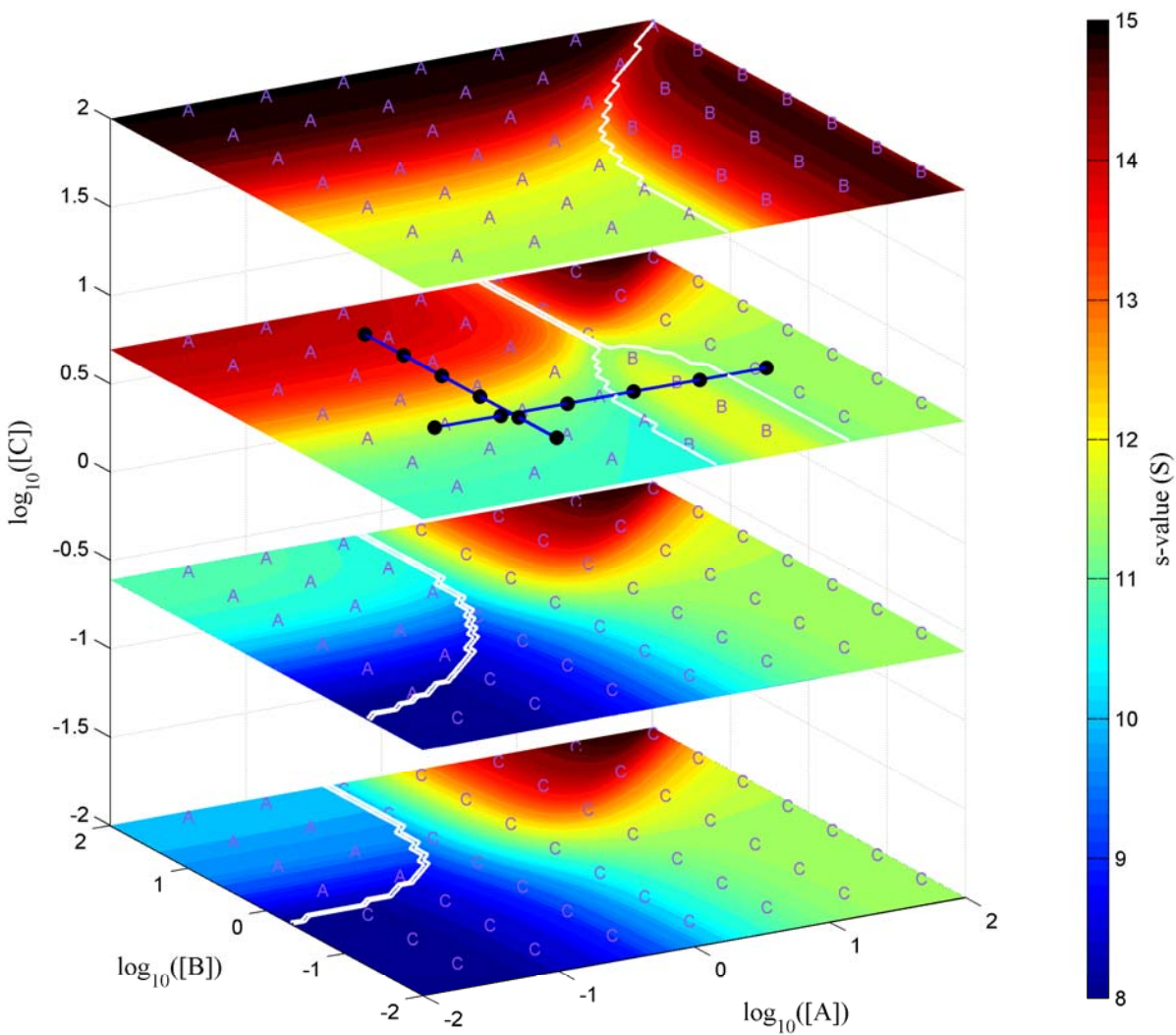

**Figure S1**: The fastest reaction boundary $s$-value can be plotted as a function of loading concentration for the same three-component system as in Fig. 1. Concentrations are plotted relative to the equilibrium dissociation constant, $K_D$, which is here equal for all interactions. Slices of the concentration space are represented as color contour plots for values of constant $\chi_{C,\text{load}}$ at 0.01, 0.24, 4.9, and 100 $K_D$. Letters indicate the component that is `dominant', i.e., solely present in the fastest boundary and governing its $s$-value. The white lines indicate the cross-section of the phase transitions shown in Fig. 3. The black circles and lines represent the trajectories for the Lamm equation simulations in Fig. 4.

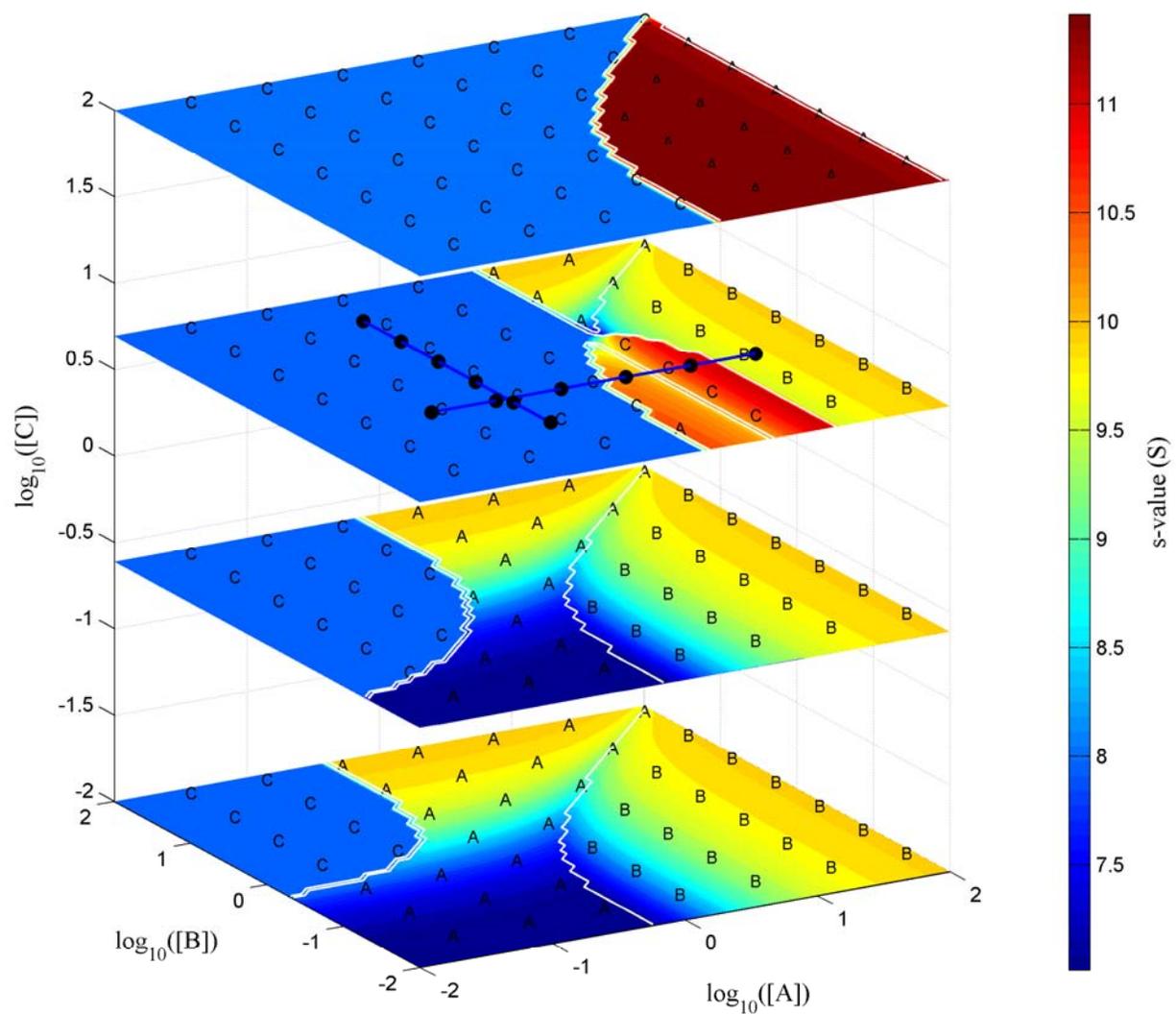

**Figure S2**: In the same representation as Fig. S1, this shows the $s$-value of the middle reaction boundary velocity in concentration space. The letters indicate the component that is dominant for the middle boundary, i.e., also contributing to the fastest reaction boundary but being excluded from the undisturbed boundary.